\input{psfig.sty} 

\documentstyle[editedvolume,psfig]{crckapb}

\begin{opening}
\title{Are Elliptical Galaxies Really Metal-Rich?}
\subtitle{And, If Not, Then What?}

\author{M. LOEWENSTEIN}
\author{R. F. MUSHOTZKY}
\institute{NASA/GSFC\\
           Code 662, Greenbelt, MD 20771, USA}

\end{opening}

\runningtitle{Elliptical Galaxy Abundances}

\begin{document}


\section{Introduction}

In elliptical galaxies, where most of the stars --
and therefore most of the heavy elements -- were formed at
an early epoch,  
the total mass,
spatial distribution, and relative abundances of metals
are intimately connected to
the galaxy formation process.

Metallicities of the stars in elliptical galaxies
can be estimated from optical broad-band
photometry and spectroscopic measurement of selected
absorption line indices.
However, the transformation from these indirect abundance indicators
to true stellar metallicites is highly non-trivial, depending on
{\it a priori} assumptions about stellar physics details and star formation
history. Since the hot
interstellar medium primarily originates from stellar mass loss,
X-ray observations provide independent and complementary abundance
information. Moreover, there are distinct advantages to using strong
X-ray emission features over weak optical absorption
features: the former are related to
actual abundances in a much more direct, model-independent, manner; and,
one can determine abundances much further out in radius. 
30-40 early-type galaxies have been observed with {\it ASCA}, 
representing the first large sample of high quality, broad-band
X-ray spectra of elliptical galaxies.
Most of these have been analyzed in detail by K. Matsushita, and we 
draw upon the results presented in her thesis,
as well as our own analysis, in this review.

\section{Abundances in Gas-Rich Ellipticals}

{\it ASCA} spectra can generally be decomposed into soft and hard
components (Matsumoto {\it et al.} 1997).
The soft component originates in the hot (0.3-1 keV) ISM, and
shows a wide range of X-ray-to-optical flux ratios and X-ray extents for
any given optical luminosity. 
The hard component
generally scales linearly 
with optical luminosity, with a relative
normalization and spectrum
consistent with  measurements of
the integrated emission from low mass X-ray binaries in spiral galaxy
bulges. 
Figure 1 (from Matsumoto {\it et al.} 1997) compares
{\it ASCA} Solid State Imaging Spectrometer (SIS) spectra
of the elliptical galaxies NGC 4365 and NGC 4636. Note that these
galaxies have similar optical luminosities and hard components;
however, the soft component in NGC 4636 is nearly two
orders of magnitude more luminous than that in NGC 4365.
Since
abundance uncertainties become large as the hard component begins
to dominate
and the equivalent widths of emission lines
are diluted (e.g., for NGC 4365), this review focuses
on observations of eighteen gas-rich ellipticals (e.g., NGC 4636)
in the {\it ASCA} archive -- 
galaxies with high
X-ray-to-optical fluxes based on
{\it ROSAT} PSPC observations and low hard component fractions
based on {\it ASCA} spectral decomposition. 
The X-ray emission extends well beyond the optical isophotes for
some of these systems, but is more compact in others.

\begin{figure}[htbp]
\centerline{
\psfig{file=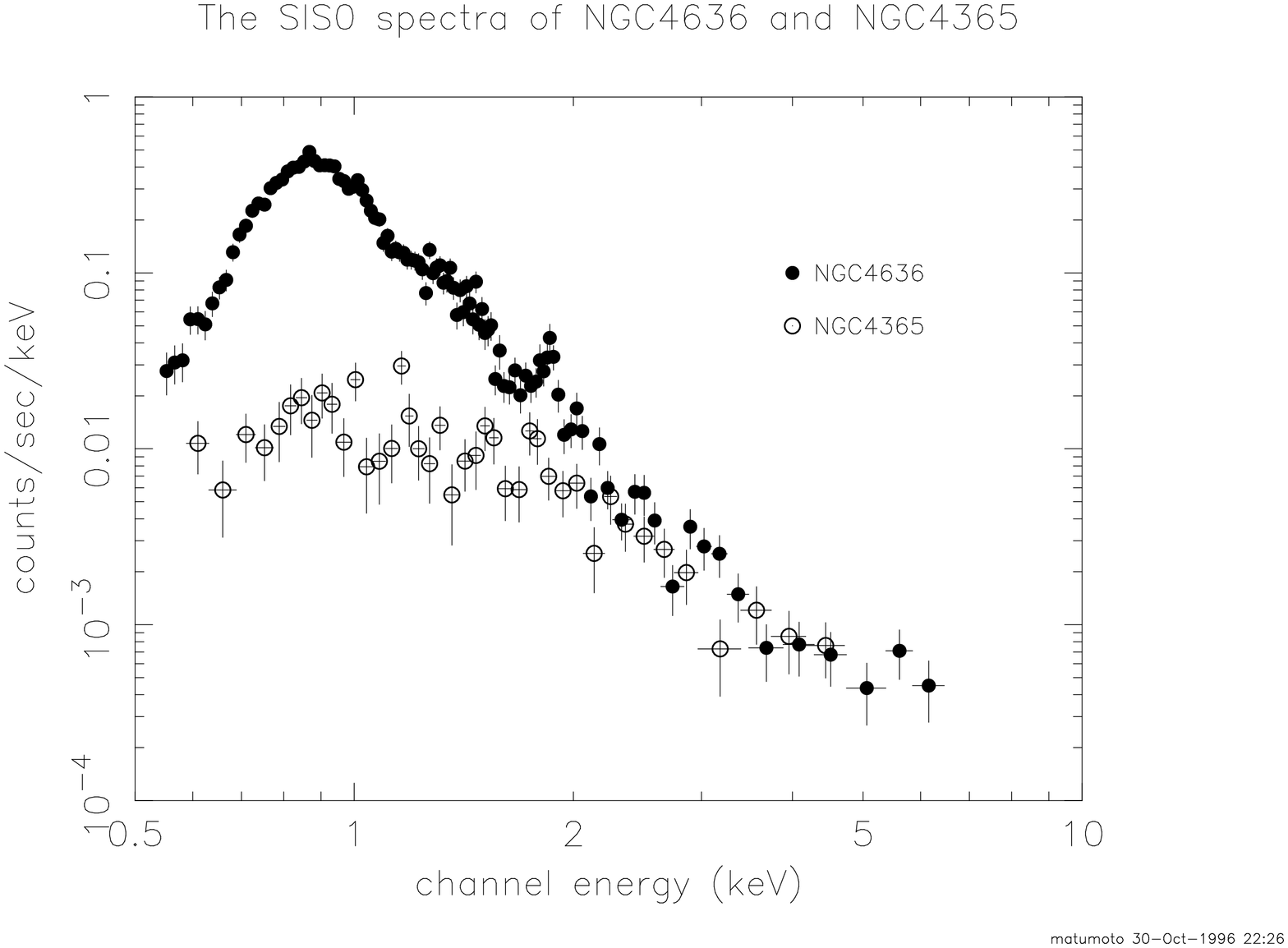,width=4.50in,height=3.60in,angle=-90}}
\caption{Co-added {\it ASCA} SIS spectra of NGC 4636 (filled circles) and 
NGC 4365 (open circles).}
\end{figure}

Figure 2 shows a plot of abundance versus temperature 
derived from {\it ASCA} spectra
extracted from the inner five optical half-light radii.
The soft component is modeled using the Raymond-Smith thermal
plasma emission code
with abundances fixed at their solar photospheric ratios. The abundances --
essentially the Fe abundance, since X-ray spectra 
at these temperatures are dominated by Fe L emission lines --
range from about one-tenth to two-thirds solar. Since it has 
generally been assumed that abundances of the mass-losing stars that
are the origin of the hot gas are supersolar,
and that Type Ia SN should further enrich the hot gas by
at least an additional two times solar, 
X-ray abundances of elliptical galaxies
are 3-30 times lower than what might
naively be expected.

\begin{figure}[htbp]
\centerline{
\psfig{file=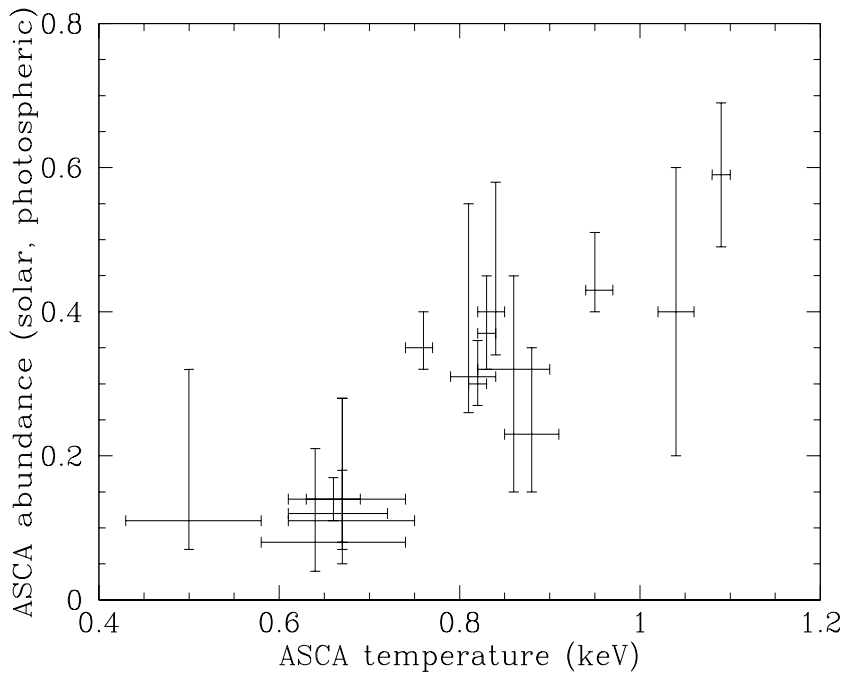,width=4.50in,height=3.60in,angle=-90}}
\caption{Hot gas metal abundance versus temperature.}
\end{figure}

\section{Reconciling Optical and X-ray Abundances}

Clearly,
either the SNIa rate is much lower than estimated,
and/or SNIa
ejecta
is not efficiently mixed into the hot ISM. Can optical
and X-ray abundances be reconciled, even if SNIa enrichment is
neglected?
It must be the case that either
(1) stellar abundances are not actually supersolar, 
(2) hot gas abundances are not actually subsolar, or
(3) X-ray and optical observations are not commensurate.
We consider these possibilities in reverse order. 

(3) We now know
that the X-ray emission from some elliptical galaxies is
much more extended than the optical light and, accounting for
the  measured
abundance distributions, that the radius containing half of the
hot gas metals is 3-10 times larger than the corresponding
radius for the stars in such systems. The hot gas metallicity 
profile generally has a negative gradient (Matsushita 1997). 
These facts indicate that
accretion of
primordial or intergalactic
material may be diluting the
abundances of the hot gas relative to the stars
(Brighenti \& Mathews 1997). 
However, if this were a dominant effect
one would expect lower average metallicities in more gas-rich galaxies --
if anything, the opposite
is observed. Although gas flows have undoubtedly
rearranged and in some cases
diluted the gas, it is probably meaningful to compare optical
and X-ray metallicities within the optical radius.

(2) The validity of the abundances obtained from single-phase
hot gas plus X-ray binaries
fits to {\it ASCA} data have been questioned on a number of fronts.
Inadequacies in the treatment of Fe L
transitions in plasma codes used for spectral fitting are an extra source
of uncertainty (Arimoto {\it et al.} 1997); however, excluding the energy
region in question in spectral fits does not systematically raise the
metallicity (Buote \& Fabian 1997). Higher abundances can also be
accommodated in more complex spectral models. In particular,
Buote and Fabian have recently found
that the best fit to {\it ASCA} data often consists of a two-temperature
plasma, with the
secondary component having a temperature of $\sim 1.5$ keV,
and that the abundances in such fits are systematically higher by
about a factor of two compared to models with a single gas phase
plus X-ray binaries.
However, there is more information in {\it ASCA}
spectra than what can be obtained from global spectral fits alone.
We have found that the temperature
obtained from the                     
He-like to H-like Si
line ratios is in excellent
agreement with the single-phase model (see
Figure 3 for an example with a typical level of agreement), 
and is not consistent with
the presence of hotter gas in the amounts suggested by
Buote and Fabian.

\begin{figure}[htbp]
\centerline{
\psfig{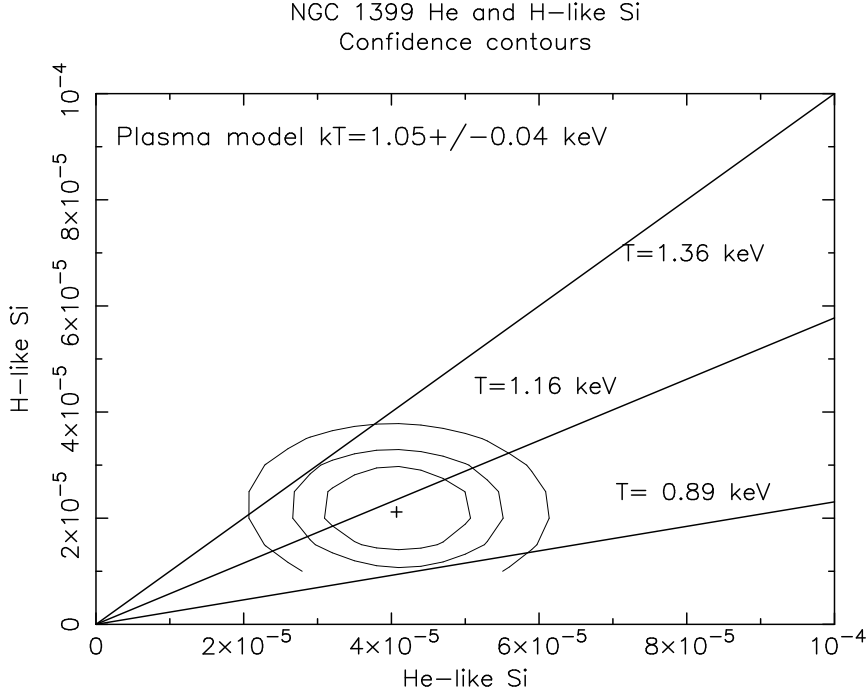}}
\caption{68, 90, and 99\% confidence contours
for H- and He-like Si line strengths (in photons cm$^{-2}$ s$^{-1}$) in
the elliptical galaxy NGC 1399. The solid lines show the ratios
expected in three single-temperature thermal plasma models.
The best-fit temperature to the spectrum derived from global
Raymond-Smith plasma model fits
is $1.05\pm 0.04$ keV (90\% confidence uncertainties).}
\end{figure}

\begin{figure}[htbp]
\centerline{
\psfig{file=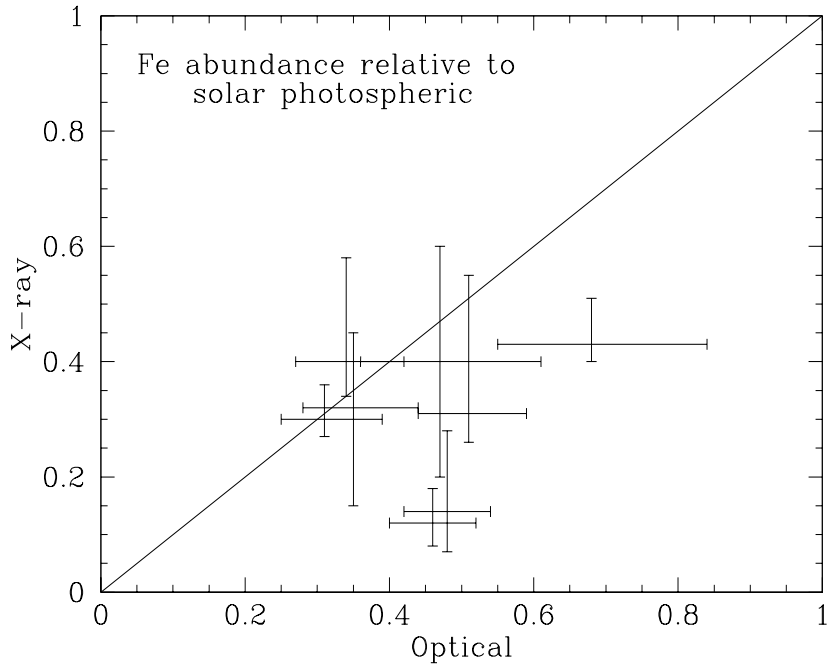,width=4.50in,height=3.60in,angle=-90}}
\caption{X-ray versus optical global iron abundance.}
\end{figure}
 
\begin{figure}[htbp]
\centerline{
\psfig{file=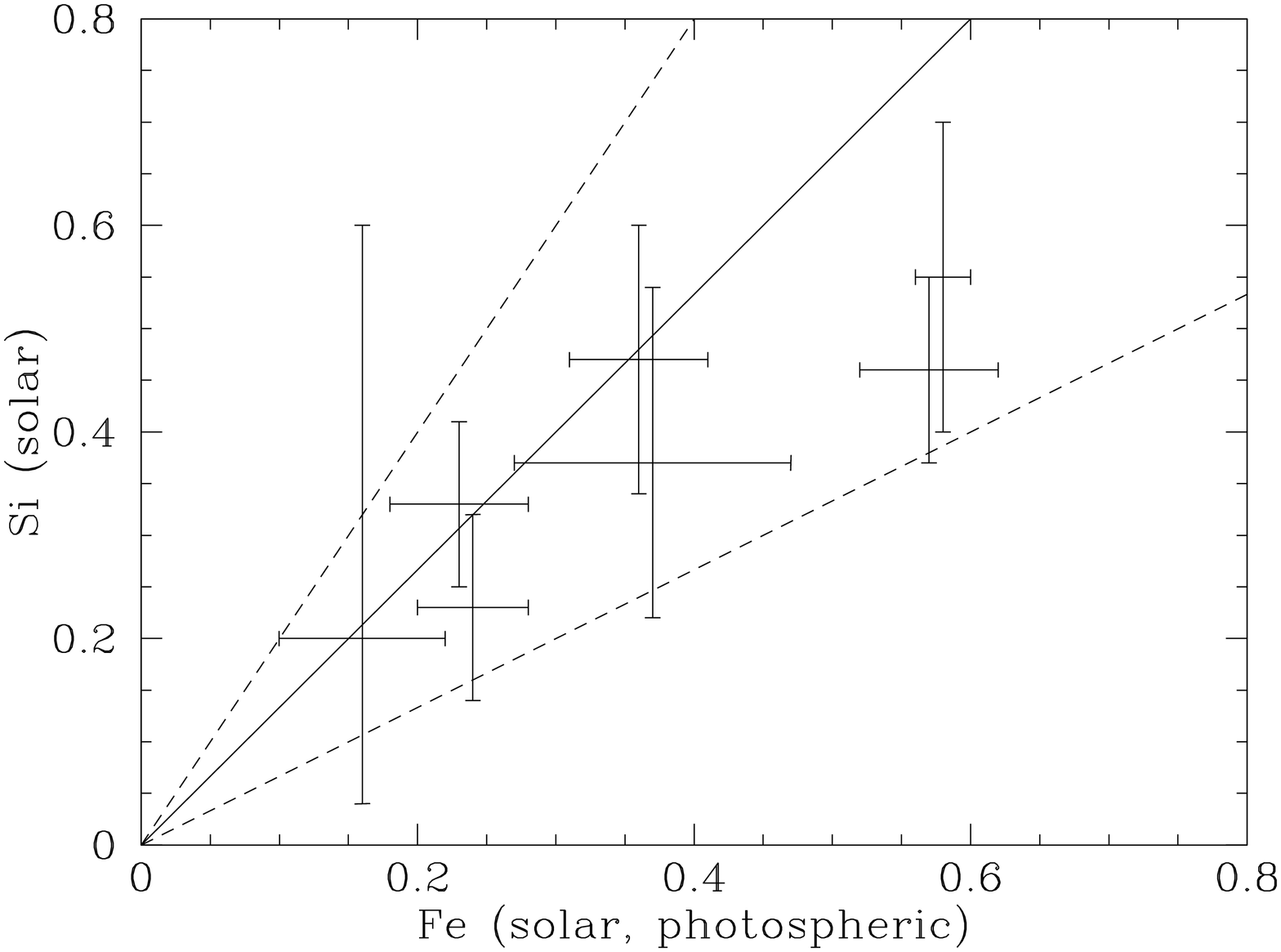,width=4.50in,height=3.60in,angle=-90}}
\caption{Si versus Fe abundance in the hot X-ray emitting gas.
The solid line denotes Si:Fe in the ratio 1:1, while
the broken lines denote the ratios 3:2 and 1:2
with respect to the (meteoritic) solar ratio.}
\end{figure}

(1) So, it seems reasonable to adopt the metallicities in the
single-phase
hot gas plus X-ray binaries fits to {\it ASCA} spectra
as measures of global interstellar Fe abundances, and sensible
to compare these with global Fe abundances of the stars.
What are the best current estimates of the latter?
The conventional wisdom that elliptical galaxies have supersolar
abundances is based on measurements of the nuclear Mg2 index. However,
when one accounts for the factor
of 2-3 overabundance of Mg with respect to Fe in the stars
(Worthey {\it et al.} 1992),
and the factor of two aperture correction (Arimoto {\it et al.} 1997)
it becomes clear that the global Fe abundances of the stars and hot gas
are not grossly discordant.  
We have compared recent estimates of the global Fe abundance using
optical data
kindly provided by S. Trager (Trager 1997) with the X-ray measurements.
The comparison for the eight galaxies present in both samples is
shown in Figure 4.
The discrepancy is generally minor, although a few galaxies
still have unaccountably low X-ray abundances.
The average optical
Fe abundances are about 0.45 solar 
(as predicted in the chemically consistent evolutionary
models of Moller {\it et al.} 1997) compared to 0.3 solar for the hot gas.

\section{Relative Abundances}

Elemental abundance ratios provide constraints on the primordial
IMF and relative numbers of Type Ia and Type II supernovae.
Given sufficient signal-to-noise, O, Mg, 
Si, S, and Fe all have prominent, distinct features
in {\it ASCA} spectra of elliptical galaxies.
In variable abundance fits of seven ellipticals, we find Si-to-Fe ratios
consistent with or perhaps somewhat less than the (meteoritic)
solar value (Figure 5). This is lower than the
Mg-to-Fe ratio derived from nuclear optical spectra, and is more in line with
what has been measured from the X-ray spectra of intergroup media.
The Si abundance provides an independent 
and robust (plasma code
uncertainties are very small for these Si lines) strong
upper limit on the effective SNIa
rate that is consistent with what is derived using Fe. The
limit of about 0.03 SNU (1 SNU = 1 supernova per century
per $10^{10}L_{B\odot}$)
is about four times lower than the
recent estimate of Cappellaro {\it et al.} (1997) for
$H_o=75$ km s$^{-1}$ Mpc$^{-1}$

\section{Implications of Low Abundances for ICM Enrichment}

We have shown that the
mass-averaged Fe abundance in an elliptical galaxy, 
based on X-ray and optical
observations, is about one-half solar. This is only slightly
higher than what is typically
measured in the intracluster medium, and yet the mass in the ICM
is generally 2-10 times greater. There is several times more Fe
(and Si, as well) in the ICM than is locked
up in stars in cluster galaxies. This implies the following.

(1) If the stellar and ICM metals come from the same SNII-enriched
proto-elliptical galaxy gas, then 50-90\% of the original galaxy mass has
been lost and a significant fraction
of the ICM is not primordial but has been ejected
from galaxies.
(2) However, the actual amount of material directly associated
with the SNII ejecta is roughly an
order of magnitude less. If there is
selective mass-loss of nearly pure SNII ejecta it is possible to
lose most of the metals without losing most of the mass.

(3) It is also possible that there
is an additional significant source of ICM enrichment, 
although there are some difficulties with such a  scenario
(Gibson and Matteucci 1997). Perhaps, alternate
sources of enrichment should be
re-examined in light of the downward revision of elliptical
galaxy abundances; although, one must bear in mind that
ICM Fe mass is highly correlated with the total luminosity
in elliptical galaxies (Arnaud {\it et al.} 1992).

\section{Concluding Remarks}

X-ray spectra of elliptical galaxies are adequately fit by models
consisting of hot gas with subsolar Fe abundance and
roughly solar Si-to-Fe ratio, plus
a hard component from an ensemble of X-ray binaries.
The consistency of the relative magnitude and spectrum
of the hard component with that expected from X-ray binaries, along with
its more compact spatial distribution supports this model over
ones where the hard component is primarily due to a hotter
gas phase. Complications in the form of an extra soft continuum
or multiple phases can be considered, but the consistency of the Si line
diagnostic and continuum temperatures demonstrates that 
the data -- at the present level of sensitivity and spectral resolution --
do not require these.
Optical and X-ray   
Fe abundance estimates are converging, although there are some
cases with anomalously low X-ray values.
Problems in the Fe L spectral region remain; however,
the main effect of improvements in atomic physics parameters is
likely to be improved spectral fits rather than a radical upward
revision in abundances.

Occam's razor would seem to demand that we provisionally
accept the reality of
low abundances in elliptical galaxies. As a result, we need to
seriously
reevaluate our notions of elliptical galaxy chemical
evolution, intracluster enrichment, and Type Ia supernova rates.

\thanks
We are grateful to Scott Trager and Kyoko Matsushita for
making results from their dissertations
available to us prior to publication.

\section{References}
\noindent 
Arimoto, N., Matsushita, K., Ishimaru, Y., Ohashi, T., \& 
Renzini, A. 1997, 

{\it ApJ}, {\bf 477}, 128

\noindent
Arnaud, M., Rothenflug, R., Boulade, O., Vigroux, L., \& Vangioni-Flan, E. 

1992, {\it A\& A}, {\bf 254}, 49

\noindent
Brighenti, F. \& Mathews, W. G. 1997, {\it ApJ}, in press

\noindent
Buote, D. A., \& Fabian, A. C. 1997, {\it MNRAS}, submitted

\noindent
Cappellaro, E., Turatto, M., Tsvetkov, D. Yu., Bartunov, O. S.,
Pollas, C., Evans, R., \& Hamuy, M. 1997, {\it A\& A}, {\bf 322}, 431

\noindent
Gibson. B. K., \& Matteucci, F. 1997, {\it ApJ}, {\bf 475}, 47

\noindent
Matsumoto, H., Koyama, K., Awaki, H., Tsuru, T., Loewenstein, M., \&
Matsushita, K. 1997, {\it ApJ}, {\bf 482}, 133

\noindent
Matsushita, K. 1997, Ph.D. thesis, University of Tokyo

\noindent
Moller, C. S., Fritze-v. Alvensleben, U. \& Fricke, K. J.
1997, {\it A\& A}, {\bf 317}, 

676

\noindent
Trager, S. C. 1997, Ph.D. thesis, University of California, Santa Cruz

\noindent
Worthey, G., Faber, S. M., \& Gonzalez, J. J. 1992, {\it ApJ}, {\bf 398}, 69

\end{document}